\documentstyle[12pt,eclepsf]{article}
\begin{document}

\begin{flushleft}
{\it Yukawa Institute Kyoto}
\end{flushleft} 

\begin{flushright}
YITP-97-66 \\
December, 1997 \\
hep-ph/9712341
\end{flushright} 
\vspace{1.0cm}

\begin{center}
\Large\bf{Scaling Distribution of Axionic Strings \\
and Estimation of Axion Density \\
from Axionic Domain Walls}
\vskip 1.0cm

\large{Michiyasu NAGASAWA}

\large\sl{Yukawa Institute for Theoretical Physics, \\
Kyoto University, Kyoto 606-01, Japan \\ \vspace{1.0cm}
nagasawa@yukawa.kyoto-u.ac.jp}
\end{center}
\vskip 1.5cm

\begin{abstract}
When the amount of the axion emitted from axionic strings
is evaluated, the number density of the source string should be
necessarily determined. Although it is the simplest assumption
that the axionic strings obey the same distribution as that of
the cosmic gauged strings, this assumption has not been completely
justified. The damping due to the surrounding plasma is estimated by
a detailed calculation of the fermion scatter from a global string
and it is shown that the friction force on the axionic strings can
be neglected before the axionic domain wall formation
in most cases, which is one of the important necessary conditions
for the scaling distribution of the strings. The energy density of
the axion produced by the collapse of the axionic domain walls is
also evaluated using the axionic string distribution, and it is found
to be larger or at least comparable to the axion density which
originates from the axionic strings.
\end{abstract}

\vfill
\begin{center}
Published in Progress of Theoretical Physics {\bf 98} (1997) 851.
\end{center}

\section{Introduction}

The dark matter problem is one of the most important cosmological
problems.\cite{Peeb} One promising candidate for cold dark matter
is the axion.\cite{PQ,axon} This is a pseudo-Nambu-Goldstone boson
associated with the Peccei-Quinn $U(1)$ symmetry breaking.\cite{PQ}
The introduction of this $U(1)$ symmetry is the most
natural solution for the strong CP problem.\cite{tHot}
Thus the existence of the axion is not only motivated by this problem
in particle physics but can also contribute to the cosmological
dark matter problem at the same time.

The axion model has two important parameters. One is the breaking energy
scale of the Peccei-Quinn symmetry, $f_a$, which determines the magnitude
of the axion mass. The other is the color anomaly, $N$, which is an
integer and equal to the number of the ground states of the effective
potential for the axion field, which corresponds to the phase of
the Peccei-Quinn field. The determination of the numerical values of
these parameters is closely related to the evolution of the topological
defects included in the axion model.

In the axion model context, at the spontaneous symmetry breaking of
the Peccei-Quinn symmetry and at the mass acquisition of the axion
around the QCD scale, topological defects are produced in the process
of cosmological evolution.\cite{topdef,VS} Since the Peccei-Quinn
symmetry is a global $U(1)$ symmetry, global strings are produced at
the scale $f_a$. These strings are axionic strings. At the QCD scale,
a $Z_N$ discrete symmetry appears through the instanton effect, and
this symmetry is broken spontaneously so that two-dimensional
topological defects stem from the axionic strings. These are axionic
domain walls and $N$ pieces of them stretch out from each axionic string.

When $N$ is greater than one, the string-wall system survives long
enough so that the energy density of the axionic domain walls dominates
the universe, and the standard cosmological evolution description should
fail, since multiple walls can stabilize the axionic string
network.\cite{RPS} Therefore the $N>1$ domain wall is not accepted
cosmologically and $N$ must be unity. When $N=1$, the axionic
domain walls disappear quickly enough not to over-close
the universe.\cite{Naga}

The Peccei-Quinn symmetry breaking scale, $f_a$, is strictly constrained
by considerations of accelerator experiments and various astrophysical
and cosmological observations. The permitted region is given by
\begin{equation}
10^{9-10}{\rm GeV}~\raisebox{-1ex}{$\stackrel{\textstyle <}
{\sim}$}~ f_a ~\raisebox{-1ex}{$\stackrel{\textstyle <}
{\sim}$}~ 10^{11-12}{\rm GeV}\ ,
\end{equation}
which is called the axion window.\cite{KT,BS} The upper bound on $f_a$
is found by the constraint that the total amount of the axion should
not be contradicted by observational facts, since the axion
contribution to the present density depends on $f_a$.
The main production mechanism of the axion is the collapse of
topological defects. The present energy density of the axion
due to axionic strings normalized by the critical density of
the universe is estimated to be\cite{BS}--\cite{HS}
\begin{equation}
\Omega_a({\rm string}) h^2 = 10^{-5 \sim -3} \left(\frac{f_a}{10^{9}
{\rm GeV}}\right)^{1.175}\ ,
\label{eq:ostring}
\end{equation}
where $h$ is the Hubble constant in units of 100\ km/sec/Mpc, and
comparable axions are supplied by the axionic domain walls as\cite{NK}
\begin{equation}
\Omega_a({\rm wall}) h^2 \simeq 10^{-3}
\left(\frac{f_a}{10^{9}{\rm GeV}}\right)^{1.175}\ .
\label{eq:owall}
\end{equation}
On the other hand, the contribution from the axion field oscillation
due to the axion field misalignment is considered to be\cite{misa,Turner}
\begin{equation}
\Omega_a({\rm mis}) h^2 = 0.4\times 10^{-3\pm 0.4}
\left(\frac{f_a}{10^{9}{\rm GeV}}\right)^{1.175}\ ,
\end{equation}
which gives a less stringent constraint on $f_a$ than Eqs.
(\ref{eq:ostring}) and (\ref{eq:owall}) do.

The above estimations suggest that in an extreme case the axion
window might be tightly closed. However, some uncertainties are not
explicitly expressed in Eq. (\ref{eq:ostring}) other than
unknown parameters in the scaling formula of number and size
distribution of the axionic strings. In addition,
the estimation (\ref{eq:owall}) is based on the assumption
that one axionic domain wall of horizon size exists
within horizon volume, which is inconsistent with the scaling
distribution of the axionic strings used in the estimation
(\ref{eq:ostring}). First of all the scale-invariant behavior
of infinitely long strings and string loops itself is confirmed only by
numerical simulations in the case of the local string. This may be
inappropriate since axionic strings are global strings without any
gauge field and have unlimitedly spread gradient energy, so that
their evolution could be modified from that of the cosmic
local strings. We pay particular attention to whether the damping of
the string motion due to the background matter can be neglected and
the scaling distribution can be realized. Although the value of
the energy which the axion emitted from the axionic string possesses
is still controversial,\cite{Davs,HS} we concentrate on the
uncertainty which depends on the nature of the distribution.

In this paper, we justify one of the necessary conditions for the scaling
behavior of the axionic string evolution. Moreover, $\Omega_a({\rm wall})$
is calculated using the same assumption of the defect distribution as
the case of $\Omega_a({\rm string})$, and the importance of the axionic
domain wall collapse is verified. In the next section, we examine
how the scattering amplitude of particles from the string is related to
the transition time of the string evolution from the friction regime to
the scaling regime. The interaction cross section between fermions and
an axionic string is calculated in $\S$3. The axion energy density
originating from the axionic domain walls is reexamined in $\S$4.
The final section is devoted to conclusions.

\section{Transition from friction regime to scaling regime}

One of the important factors for the establishment of the scaling
distribution of axionic strings is that the damping force exerted on them
by the radiation background should be negligible so that they move freely
by the time of the axionic domain wall formation, $t_1$. The temperature
at $t_1$, $T_1$, is derived from the condition that the axion mass grows
greater than the expansion rate of the universe, and the dynamics of
the wall are determined by its tension using the temperature dependence
of the axion mass,\cite{Turner}
\begin{equation}
m_a\left(T\right)\simeq 0.1m_a\left(T=0\right)\left(
\frac{\Lambda_{\rm QCD}}{T}\right)^{3.7}\ ,
\end{equation}
where $\Lambda_{\rm QCD}$ is the QCD scale.
The expression of $T_1$ is written as\cite{NK}
\begin{eqnarray}
T_1 &=& 3.56~{\rm GeV} \left(\frac{f_a}{10^9{\rm GeV}}\right)^{-0.175}
\nonumber \\ & & \times
\left(\frac{\cal N}{72.25}\right)^{-0.0877}
\left(\frac{\Lambda_{\rm QCD}}{200{\rm MeV}}\right)^{-0.65}
\left(\frac{m_a^0}{6.2\times10^{-3}{\rm eV}}\right)^{0.175}\ ,
\end{eqnarray}
where ${\cal N}=289/4=72.25$ is the relativistic degrees of freedom
at $T\sim 1$ GeV, and $m_a^0$ is the zero temperature axion mass
when $f_a=10^9$ GeV.

Conventional cosmic strings which are suitable for the initial
perturbation seeds of the cosmological structures exist in the scaling
evolution regime before $t_1$. Their energy scale, however, is typically
equal to the grand unification scale,\cite{VS} and it is much higher than
that of the axionic string formation, $f_a$. Then it might be possible that
the frictional force on the axionic strings remains dominant until $t_1$.

Actually, the temperature of the transition from the friction regime to
the scale-invariant regime, $T_*$, is written as\cite{VS}
\begin{equation}
T_* \sim 30~{\rm GeV} \left( \frac{f_a}{10^9{\rm GeV}}\right)^2
\left(\frac{\cal N}{72.25}\right)^{-1/2}\ ,
\label{eq:ttime}
\end{equation}
which is a rough estimation. It might not be a trivial matter whether
$T_*$ is well above $T_1$ and the scaling assumption holds during
the entire period of the axion emission from the axionic strings.
The longer the friction regime of the axionic strings lasts,
the smaller the contribution from the axionic string loops becomes,
since in that case all the energy which the axionic strings have \
may not be converted to axions. Then the number of the axions produced
should decrease and the allowed region of the axion model parameter,
$f_a$, could be extended. For example, if the loop formation is
suppressed and only the infinite strings exist, the expected
$\Omega_a({\rm string})$ would be even less than one tenth of
the value in Eq. (\ref{eq:ostring}), since the axion energy density
from the axionic string loops is much larger than that from
the infinite axionic strings.\cite{BS} However, the most dominant
contribution from the axionic strings comes from the low temperature
phase, that is, the time just before axionic domain wall generation,
so that the confirmation that $T_*$ cannot be smaller than $T_1$ would
be sufficient for the practical estimation of the axion density.

The formula (\ref{eq:ttime}) is derived by the condition that
the tension of the axionic strings, $F_t$, is balanced with
the friction due to the background plasma in the universe, $F_f$.
The former is written using the line energy density of the string,
$\mu_s$, and the curvature radius of the string, $R$, by
\begin{equation}
F_t=\frac{\mu_s}{R}\ .
\end{equation}
The energy density per unit length of the global string is expressed by
\begin{equation}
\mu_s \simeq f_a^2+2\pi f_a^2 \ln \frac{\xi_s}{\delta_s}\ ,
\end{equation}
where $\delta_s \sim f_a^{-1}$ is the width scale of the string
core, and $\xi_s$ is the cutoff scale, for example, the mean
separation between strings. In calculating the numerical value of
the formula (\ref{eq:ttime}), we have substituted $\sim 60$ as
$\ln (\xi_s/\delta_s)$, where the parameters are
$\delta_s^{-1}\sim 10^{9-12}$ GeV and $\xi_s\sim 2t_1$, which is
the horizon scale at $t_1$, that is, the maximum scale of
the axionic string length.

The frictional force per unit string length is proportional to
the scattering cross section of interacting particles from
the axionic string, $\sigma_s$. With the conventional formula for
the local strings, $\sigma_s$ is proportional to the inverse of $p$,
the momentum of the particle :\cite{Ever}
\begin{equation}
\sigma_s=\frac{C_s}{p}\ ,
\label{eq:lsigma}
\end{equation}
where $C_s$ is a correction factor which we will calculate in the
next section. Using the number density of the particles which interact
with the axionic string, $n$, the frictional force can be written as
\begin{equation}
F_f=\int dp~n\left(p\right)~pv_s~\sigma_s\ , \label{eq:ff}
\end{equation}
where $v_s$ is the velocity of the string motion and the momentum
transfer is evaluated as $pv_s$.

Although the scattering cross section may take various values
depending on the species of particle, we temporarily assume that
$\sigma_s$ is common to all. Then by equating $F_t$ with $F_f$ and
assuming that $R=v_st$, we obtain the relation
\begin{equation}
R\propto {\mu_s}^{1/2} {C_s}^{-1/2} T^{-1/2}t\ . \label{eq:rpro}
\end{equation}
After substitution of the usual relation between $t$ and $T$ in
the standard cosmology into Eq. (\ref{eq:rpro}), and using
the relation $R=2t$ (that is, the length of the string equals
the horizon length), we can show
\begin{equation}
T_*\propto \mu_s {C_s}^{-1}\ .
\label{eq:tpropo}
\end{equation}
One point we would like to note is that the temperature $T_*$ is
in inverse proportion to the correction factor of $\sigma_s$.
In addition, for axionic strings of smaller $\mu_s$ than
that used in the estimation (\ref{eq:ttime}) due to some kinds of
refinement, the factor $\mu_s/f_a^2$ can approach unity,
so that $T_*$ should be much lower.

The evolution of axionic strings may be controlled by
the frictional force rather than the string tension even around $T_1$
if $C_s$ is sufficiently large. When we employ a simple geometric
cross section, $C_s$ can be estimated to be $\pi$. Although we can
say this is not disastrous to the scaling approximation,
the difference between $T_1$ and $T_*$ is not enormous.
The exact value of $C_s$ should be evaluated by analysis of
the interaction between an axionic string and fermions.
The scattering problem of the particles from the string was treated
for gauged strings,\cite{loss} the electroweak strings\cite{ewss}
and the global strings.\cite{glss} However, the interaction of
an axionic string with fermions, which is necessary for the calculation
of (\ref{eq:lsigma}), has not been considered in past work.
Particularly, consideration of the case in which the coupling constant
is an integer, which is the case in the usual axion model,
is missing in the literature.
Although the momentum dependence may be predicted as
$C_s \propto \ln^{-2}p\delta_s$,\cite{Ever} the numerical
factor would be known only after actual calculation.

\section{Fermion scatter from an axionic string}

In this section, the scattering cross section of relativistic
fermions from a static, straight, infinitely long axionic string
$\sigma_s$ is computed. Since the local cosmic string has been
the main object of study in this field, a systematic picture for
the global string is lacking. Hence we would like to show not only
the result but also the calculation procedure briefly.

Here we consider
the derivative coupling of the axion field with fermions,\cite{KT,Kim}
where the non-zero vacuum expectation value of the Peccei-Quinn field
does not influence their mass.
The interaction Lagrangian is expressed by
\begin{equation}
{\cal L}_f =g_f \frac{\partial_\mu A}{f_a}\overline{\Psi}
\gamma^\mu\Psi\ ,
\end{equation}
where $g_f$ is a coupling constant, $A$ is the axion field, and
$\Psi$ is a fermion spinor which we regard as massless for simplicity.

In the background, where a global axionic string exists, $A$ is
spatially dependent. As a result, the Dirac equation is altered
to the formula
\begin{equation}
\gamma^\mu \left( i\partial_\mu +g_f \frac{\partial_\mu A}{f_a}
\right)\Psi=0\ .
\end{equation}
When the gamma matrix representation as
\begin{equation}
\gamma^0=\left(
\begin{array}{cc}
0 & {\bf 1} \\ {\bf 1} & 0
\end{array}
\right)\ ,\quad \gamma^k=\left(
\begin{array}{cc}
0 & -\sigma^k \\ \sigma^k & 0
\end{array}
\right),
\end{equation}
where $k=1,2,3$ is employed, the equations for the upper
and lower two components of the spinor are separated as
\begin{equation}
\left(
\begin{array}{l}
0\quad\quad 0\quad\quad i\left(\partial_t-\partial_z\right)
+\frac{g_f}{f_a}\left(\partial_t-\partial_z\right) A\quad\quad
-ie^{-i\theta}\partial_--\frac{g_f}{f_a}e^{-i\theta}\partial_- A \\
0\quad\quad 0\quad\quad -ie^{i\theta}\partial_+-\frac{g_f}{f_a}
e^{i\theta}\partial_+ A\quad\quad i\left(\partial_t+\partial_z\right)
+\frac{g_f}{f_a}\left(\partial_t+\partial_z\right) A \\
i\left(\partial_t+\partial_z\right)+\frac{g_f}{f_a}\left(\partial_t
+\partial_z\right)A\quad\quad ie^{-i\theta}\partial_-+\frac{g_f}{f_a}
e^{-i\theta}\partial_- A\quad\quad 0\quad\quad 0 \\
ie^{i\theta}\partial_++\frac{g_f}{f_a}e^{i\theta}\partial_+ A
\quad\quad i\left(\partial_t-\partial_z\right)+\frac{g_f}{f_a}
\left(\partial_t-\partial_z\right) A\quad\quad 0\quad\quad 0
\end{array}
\right)\Psi=0\ ,
\end{equation}
\begin{equation}
\partial_\pm \equiv \partial_r\pm i\frac{\partial_\theta}{r}\ ,
\end{equation}
where $(r, \theta, z)$ are components of the cylindrical coordinate
whose $z$-axis coincides with the string axis. In the case of
the static, straight, axial symmetric axionic string,
$A$ depends only on the azimuthal angle $\theta$.
Hence $\partial_t A$, $\partial_r A$ and $\partial_z A$ must
be zero, which are substituted hereafter.

In order to solve the scattering problem, the Dirac spinor is
decomposed into eigenstates of the angular momentum around
the string axis as usual :
\begin{equation}
\Psi=\sum_{j=-\infty}^{+\infty}\left(
\begin{array}{c}
\psi_1^j(r) \\ i\psi_2^j(r)e^{i\theta} \\
\psi_3^j(r) \\ i\psi_4^j(r)e^{i\theta}
\end{array}
\right) e^{ij\theta +ip_z z-i\omega t}\ ,
\end{equation}
where $\omega$ is the total energy of the fermion, and $p_z$ is
the $z$-component of the fermion momentum. The equations for
each decomposed spinor are simplified to
\begin{eqnarray}
& & \left( \frac{d}{dr}-\frac{j}{r}+\frac{g_f}{f_a}
\frac{\partial_\theta}{r} A
\right)\psi_1^j +\left( \omega+p_z \right) \psi_2^j =0\ ,
\label{eq:right1} \\
& & \left( \frac{d}{dr}+\frac{j+1}{r}-\frac{g_f}{f_a}
\frac{\partial_\theta}{r} A
\right)\psi_2^j -\left( \omega-p_z \right) \psi_1^j =0
\label{eq:right2}
\end{eqnarray}
for right-handed fermions and
\begin{eqnarray}
& & \left( \frac{d}{dr}-\frac{j}{r}+\frac{g_f}{f_a}
\frac{\partial_\theta}{r} A
\right)\psi_3^j -\left( \omega-p_z \right) \psi_4^j =0\ , \\
& & \left( \frac{d}{dr}+\frac{j+1}{r}-\frac{g_f}{f_a}
\frac{\partial_\theta}{r} A
\right)\psi_4^j +\left( \omega+p_z \right) \psi_3^j =0
\end{eqnarray}
for left-handed ones. Hereafter we consider the case of
right-handed components, since the calculation for
left-handed components is almost identical,
and the resulting cross section can be applied to
the left-handed case.

In a region sufficiently far from the axionic string,
the axion field takes the value of the solution for the static,
straight, infinitely long and axially symmetric string. That is,
\begin{equation}
A=f_a\theta
\end{equation}
for the string whose winding number is one. Then the spatial
derivative of $A$ is expressed by
\begin{equation}
\partial_k A=f_a\left( 0,\ \frac{1}{r},\ 0\right)
\end{equation}
for the cylindrical coordinate which we adopt.
Then the differential equations (\ref{eq:right1}) and
(\ref{eq:right2}) take the form of Bessel equations as
\begin{eqnarray}
& &\left(\omega^2-p_z^2\right)\psi_1+\left(\frac{d^2}{dr^2}+\frac{1}{r}
\frac{d}{dr}-\frac{\left(j-g_f\right)^2}{r^2}\right)\psi_1=0\ ,\\
& &\left(\omega^2-p_z^2\right)\psi_2+\left(\frac{d^2}{dr^2}+\frac{1}{r}
\frac{d}{dr}-\frac{\left(j+1-g_f\right)^2}{r^2}\right)\psi_2=0\ ,
\end{eqnarray}
and the solutions for $\psi_1$ and $\psi_2$ are
written by the Bessel functions as
\begin{equation}
\left(
\begin{array}{c}
\psi_1^j(r) \\
\psi_2^j(r)
\end{array}
\right)=\left(
\begin{array}{c}
a_j J_{\nu}\left( \tilde{r}\right)+b_j J_{-\nu}\left( \tilde{r}\right) \\
D_2 a_j J_{\nu+1}\left( \tilde{r}\right)-D_2 b_j J_{-\nu-1}
\left( \tilde{r}\right)
\end{array}
\right)\ ,
\end{equation}
when $g_f$ is not an integer and
\begin{equation}
\left(
\begin{array}{c}
\psi_1^j(r) \\
\psi_2^j(r)
\end{array}
\right)=\left(
\begin{array}{c}
a_j J_{\nu}\left( \tilde{r}\right)+b_j N_{\nu}\left( \tilde{r}\right) \\
D_2 a_j J_{\nu+1}\left( \tilde{r}\right)+D_2 b_j N_{\nu+1}
\left( \tilde{r}\right)
\end{array}
\right)\ ,
\end{equation}
when $g_f$ is an integer, where the argument of the above functions,
$\tilde{r}$, is defined by
\begin{equation}
\tilde{r}\equiv \sqrt{\omega^2-p_z^2}~r\ ,
\end{equation}
the subscript, $\nu$, is defined by
\begin{equation}
\nu \equiv j-g_f\ ,
\end{equation}
$a_j$ and $b_j$ are arbitrary constants, and
\begin{equation}
D_2\equiv \sqrt{\frac{\omega-p_z}{\omega+p_z}}\ .
\end{equation}

We consider the incoming plain wave as the boundary condition at
the region infinitely far from the axionic string for the purpose
of calculating the scattering cross section. The combination
of the plain wave which proceeds in the direction of $x$-axis
and the scattered cylindrical wave is written as
\begin{equation}
\left(
\begin{array}{c}
\psi_1^j \\
\psi_2^j
\end{array}
\right)_{r\to\infty}=\left(
\begin{array}{c}
i^jJ_j\left( \tilde{r}\right) +f_j
\frac{e^{i\tilde{r}}}{\sqrt{r}} \\
D_2i^jJ_{j+1}\left( \tilde{r}\right)
+\frac{1}{i}D_2f_j\frac{e^{i\tilde{r}}}{\sqrt{r}}
\end{array}
\right)
\end{equation}
for the case in which the plain wave comes from negative infinity and
\begin{equation}
\left(
\begin{array}{c}
\psi_1^j \\
\psi_2^j
\end{array}
\right)_{r\to\infty}=\left(
\begin{array}{c}
\left( -i\right)^jJ_j\left( \tilde{r}\right) +f_j
\frac{e^{i\tilde{r}}}{\sqrt{r}} \\
D_2\left( -i\right)^jJ_{j+1}\left( \tilde{r}\right)
+\frac{1}{i}D_2f_j\frac{e^{i\tilde{r}}}{\sqrt{r}}
\end{array}
\right)
\end{equation}
for that when the plain wave comes from positive infinity.
We can obtain the scattering cross section of the fermion
from the string per unit string length, $\sigma_s$, as
\begin{equation}
\frac{d\sigma_s}{d\theta}=\sum_{j=-\infty}^{+\infty}\left| f_j\right|^2\ ,
\end{equation}
using the coefficients of the scattered wave.

In order to evaluate $f_j$, we must determine the inside
boundary condition at the string core. For this purpose we employ
the step-function-like configuration at $r=\delta_s$. That is,
the axion field takes a trivial value in the interior of the string
core, $A=0$, and assumes the value corresponding to distances
infinitely far from the string in
the exterior of the string core. Under the condition that
the spinors are regular at the origin, $r=0$, the solutions
inside the string core are written as
\begin{equation}
\left(
\begin{array}{c}
\psi_1^j(r) \\
\psi_2^j(r)
\end{array}
\right)=\left(
\begin{array}{c}
c_j J_{j}\left( \tilde{r}\right) \\
D_2 c_j J_{j+1}\left( \tilde{r}\right)
\end{array}
\right)\ ,
\end{equation}
where $c_j$ is an arbitrary constant. Then the coefficient $f_j$
can be determined by the junction condition between the internal
solution and the external solution at $r=\delta_s$. The results are
summarized as follows (where $p_{\perp}\equiv \sqrt{\omega^2-p_z^2}$
is the fermion momentum perpendicular to the string axis and
$\tilde{r}_s \equiv \sqrt{\omega^2-p_z^2}\delta_s$).
The solutions in the case that $g_f$ is not an integer are
\begin{eqnarray}
f_j&=&\frac{1}{\sqrt{2\pi p_\perp i}}\left\{e^{-i\frac{\nu}{2}\pi}
\left(e^{i\nu\pi}-e^{-i\nu\pi}\right)b_j-1+\left(-1\right)^j
e^{-i\nu\pi}\right\}\ , \\
b_j&=&\left[\left(-1\right)^je^{-i\frac{\nu}{2}\pi}\left\{J_{j+1}
\left(\tilde{r}_s\right)J_{\nu}\left(\tilde{r}_s\right)-J_j
\left(\tilde{r}_s\right)J_{\nu+1}\left(\tilde{r}_s\right)\right\}
\right] / B_j
\end{eqnarray}
for the $+x$ oriented incoming wave and
\begin{eqnarray}
f_j&=&\frac{1}{\sqrt{2\pi p_\perp i}}\left\{e^{-i\frac{\nu}{2}\pi}
\left(e^{i\nu\pi}-e^{-i\nu\pi}\right)b_j-\left(-1\right)^j+
e^{-i\nu\pi}\right\}\ , \\
b_j&=&\left[e^{-i\frac{\nu}{2}\pi}\left\{J_{j+1}
\left(\tilde{r}_s\right)J_{\nu}\left(\tilde{r}_s\right)-J_j
\left(\tilde{r}_s\right)J_{\nu+1}\left(\tilde{r}_s\right)\right\}
\right] / B_j
\end{eqnarray}
for the $-x$ oriented incoming wave, where
\begin{eqnarray}
B_j&\equiv&J_{j+1}\left(\tilde{r}_s\right)\left\{e^{-i\nu\pi}J_{\nu}
\left(\tilde{r}_s\right)-J_{-\nu}\left(\tilde{r}_s\right)\right\}
\nonumber \\
& & -J_j\left(\tilde{r}_s\right)\left\{e^{-i\nu\pi}J_{\nu+1}
\left(\tilde{r}_s\right)+J_{-\nu-1}\left(\tilde{r}_s\right)\right\}\ .
\end{eqnarray}
Those in the case that $g_f$ is an integer are
\begin{eqnarray}
f_j&=&\frac{1}{\sqrt{2\pi p_\perp i}}\left\{-\frac{i}{2}
e^{-i\frac{\nu}{2}\pi}b_j-1+\left(-1\right)^j
e^{-i\nu\pi}\right\}\ , \\
b_j&=&\frac{\left(-1\right)^je^{-i\frac{\nu}{2}\pi}\left\{J_{j+1}
\left(\tilde{r}_s\right)J_{\nu}\left(\tilde{r}_s\right)-J_j
\left(\tilde{r}_s\right)J_{\nu+1}\left(\tilde{r}_s\right)\right\}}
{J_{j+1}\left(\tilde{r}_s\right)\left\{J_{\nu}\left(\tilde{r}_s\right)
-N_{\nu}\left(\tilde{r}_s\right)\right\}-J_j\left(\tilde{r}_s\right)
\left\{J_{\nu+1}\left(\tilde{r}_s\right)-N_{\nu+1}
\left(\tilde{r}_s\right)\right\}}
\end{eqnarray}
for the $+x$ oriented incoming wave and
\begin{eqnarray}
f_j&=&\frac{1}{\sqrt{2\pi p_\perp i}}\left\{-\frac{i}{2}
e^{-i\frac{\nu}{2}\pi}b_j-\left(-1\right)^j+
e^{-i\nu\pi}\right\}\ , \\
b_j&=&\frac{e^{-i\frac{\nu}{2}\pi}\left\{J_{j+1}
\left(\tilde{r}_s\right)J_{\nu}\left(\tilde{r}_s\right)-J_j
\left(\tilde{r}_s\right)J_{\nu+1}\left(\tilde{r}_s\right)\right\}}
{J_{j+1}\left(\tilde{r}_s\right)\left\{J_{\nu}\left(\tilde{r}_s\right)
-N_{\nu}\left(\tilde{r}_s\right)\right\}-J_j\left(\tilde{r}_s\right)
\left\{J_{\nu+1}\left(\tilde{r}_s\right)-N_{\nu+1}
\left(\tilde{r}_s\right)\right\}}
\end{eqnarray}
for the $-x$ oriented incoming wave.

Finally when $p_\perp \ll \delta_s^{-1}$, which holds in a
universe whose temperature is sufficiently lower than $f_a$,
the scattering cross section can be deduced as
\begin{equation}
\sigma_s\approx \frac{2\sin^2\left(g_f\pi\right)}{p_\perp}
\end{equation}
for a non-integer $g_f$, which has the same form as the Aharonov-Bohm
cross section,\cite{loss} and
\begin{equation}
\sigma_s\approx
\frac{\pi^2}{8p_\perp}\frac{1}{\ln^2
\left(\frac{p_\perp \delta_s}{2}\right)}
\end{equation}
for an integer $g_f$, which has a form similar to Everett's
cross section.\cite{Ever} The important features that $\sigma_s$
is proportional to the inverse of $p_{\perp}$, the logarithmic
dependence on $p_\perp \delta_s$ appears when the coupling charge
is an integer, and the sine of the charge itself is multiplied
in the contrary case, resemble very much that of the electroweak
string.\cite{ewss} When the mass of the fermions is included,
the helicity flip cross section and the helicity conserving cross
section will be defined as is the electroweak string case,
and the essential properties of the total cross section
would not be changed.

Although the above result shows that $\sigma_s$ is proportional to
$p_\perp$ rather than the total momentum, the calculation of $T_*$,
the temperature when the friction regime finishes, should not be
modified, since the momentum transfer in Eq. (\ref{eq:ff}) is
replaced by $p_\perp v_s$, and the subsequent procedure is identical.
The correction factor, $C_s$, which is determined in
Eq. (\ref{eq:lsigma}) is equal to
\begin{equation}
C_s=\left\{
\begin{array}{ll}
2\sin^2\left(g_f\pi\right) \quad & ;~g_f \notin {\bf Z}\ , \\
\frac{\pi^2}{8}\frac{1}{\ln^2\left(\frac{p_\perp \delta_s}
{2}\right)} \quad & ;~g_f \in {\bf Z}\ .
\end{array}
\right.
\end{equation}

When $p_\perp \delta_s \ll 1$, $C_s$ decreases extremely rapidly.
Thus when $g_f$ is an integer or a sufficiently small non-integer,
the interaction of the axionic strings with the background fermions
is fairly strongly suppressed, so that the frictional force becomes
insignificant much earlier than the time of the axionic domain wall
formation, $t_1$. Even in a parameter range in which $g_f$ is
nearly a half-integer, $C_s$ is at most approximately 2.
The correction factor cannot be much larger than unity,
and the estimation of $T_*$ in the representation
(\ref{eq:ttime}) is not completely unreasonable.
Hence $T_* > T_1$ is shown to be a good approximation.

\section{Revised estimation of axion density}

Now the wall number density can be explicitly calculated using
the distribution formula of the axionic strings. Once the scaling
approximation is justified, we can utilize numerical simulations
of strings in which the frictional force they experience is not included.
Then the numerical values of the parameters in the string distribution
formula enable the quantitative estimation of the axion density from
the axionic domain walls. Until now, the wall contribution has been
neglected,\cite{BS} or its treatment is based on a naive
simplification.\cite{NK} As we will see, this contribution is not
minor compared with the string contribution.

In the framework of the one-scale model, the smoothing of the small
scale structure of infinitely long strings which supplies additional
string loops, and the energy loss of loops occur continuously.
The resulting energy density of the strings has certain formulae,
although some model parameters are left inexactly determined.\cite{VSa}

First, the energy density of infinitely long axionic strings
at $t_1$ is written as
\begin{equation}
\rho_s^\infty\left(t_1\right)=\mu_s \frac{\zeta}{t_1^2}\ ,
\label{eq:rhols}
\end{equation}
where $\zeta$ is the inverse squared correlation scale, since we
can expect that the large-scale string network should be formulated
by Brownian random walks. Its numerical value has been estimated
to be $\zeta\sim 13$ by numerical simulations of cosmic
strings.\cite{VSa,BBAS}

On the other hand, in order to determine
the size and number distribution for the loops, we must acquire
the complete formula of that for the loops created from the long
strings. However, such a determination is beyond current computational
power. Therefore we utilize the simple assumption that all loops
have the same relative size to the horizon length. Then the energy
density contribution from the axionic string loops during
the radiation dominated era in the interval from $l$ to $l+dl$ is
expressed by
\begin{equation}
\rho_s^{\rm loop}\left(t_1\right)dl=\mu_s l\frac{\nu_a}{t_1^{3/2}
\left(l+\kappa t_1\right)^{5/2}}dl\ , \label{eq:rholp}
\end{equation}
which is valid for the range of length $l \le \alpha t_1$,
where $\alpha t$ is the assumed loop creation size at $t$.
Although the lower bound of the region where the above formula is
effective should be defined by the minimum size of the loop, we set
it to zero. This does not affect the final conclusion.
The normalization factor, $\nu_a$, is defined by
\begin{equation}
\nu_a =C_a\zeta\alpha^{1/2}\ ,
\end{equation}
where the parameter $C_a =(1-\left<v_s^2\right>)$ is defined
by the velocity of the strings. Using the value of
the velocity dispersion of the cosmic string motion,
$\left<v_s^2\right>,$ obtained by string simulations, $C_a \sim 0.4$
can be deduced.\cite{VSa,KBAT}
The radiation back-reaction scale $\kappa t$ is derived as
\begin{equation}
\kappa =\frac{\Gamma_a}{2\pi}\ln^{-1}\frac{t_1}{\delta_s}\ ,
\end{equation}
which defines the energy loss rate of a string loop and suggests
that the lifetime of a loop created at $t$ should be approximately
$\alpha/\kappa\times t$.\cite{VSa,VV} The parameter in this formula
is the efficiency factor of the axion emission power from the axionic
string loops, $\Gamma_a$, and it should be determined by
numerical simulations of global strings as
$\Gamma_a \sim 65$.\cite{VSa,AS}
The numerical value of $\kappa$ ranges from $0.15$ to $0.17$,
according to the variation of $t_1$ due to $f_a$.

\begin{figure}
\epsfile{file=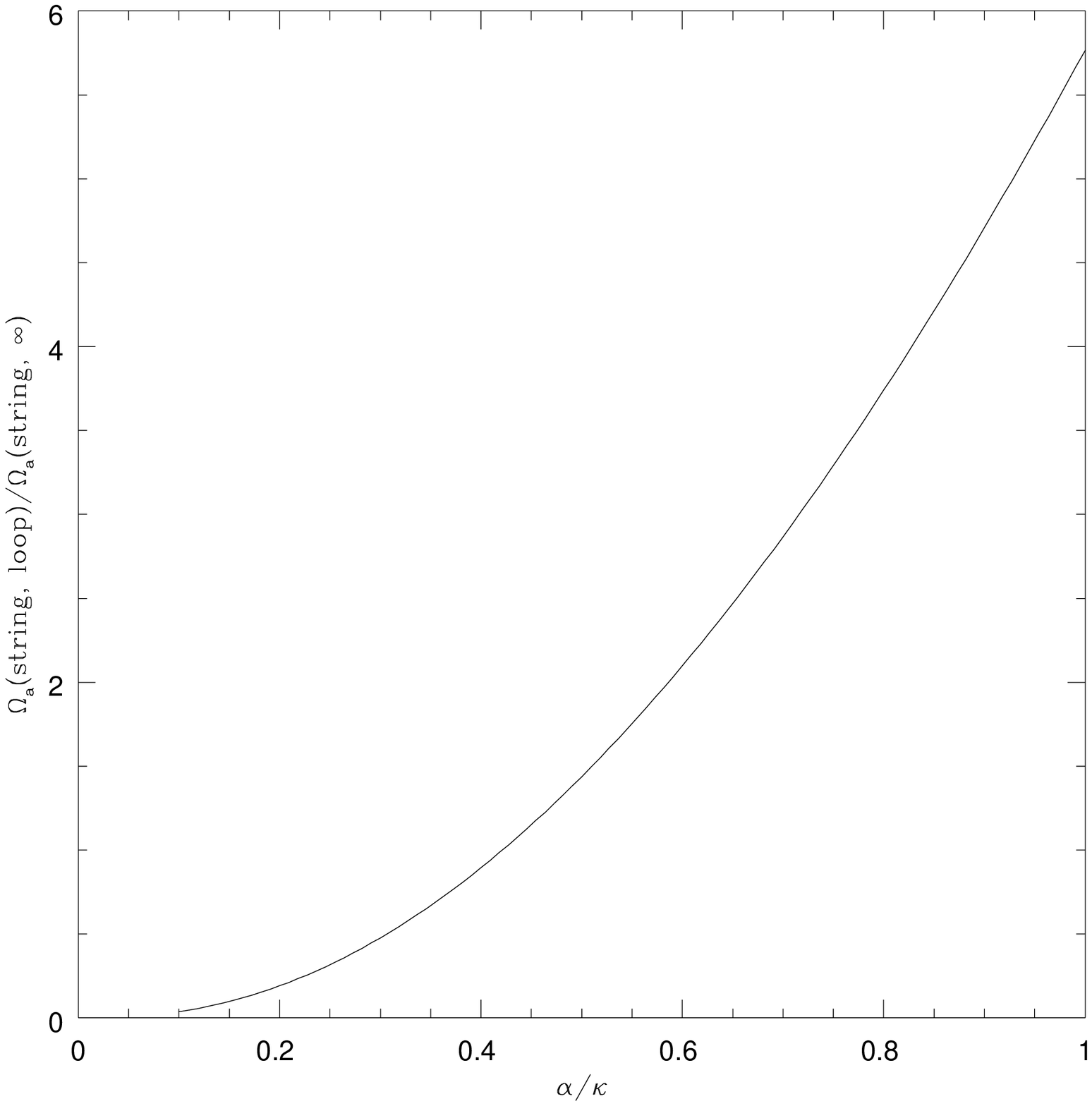,width=15.0cm}
\caption{The ratio of the energy density of the axions emitted from
the axionic string loops to that from the long axionic strings
is exhibited. The horizontal axis represents the ratio of the loop
creation size to the radiation back-reaction scale.}
\label{fig:ostring}
\end{figure}

Before we proceed to the calculation of $\Omega_a({\rm wall})$,
let us review the result for $\Omega_a({\rm string})$ briefly.
The following numerical value is basically quoted from
Ref. \cite{BS}, but the normalization is different;
the expression of $\nu_a$ is replaced by that in Ref. \cite{VSa}
and the estimation is carried out at $t_1$, not $t_w$ when
the line energy of the horizon size loop becomes equivalent to
the surface energy of the domain wall surrounded by this loop :
\begin{equation}
\sigma_w t_w^2=\mu_s t_w\ .
\end{equation}
Here the surface energy density of the axionic domain wall is
written as\cite{topdef}
\begin{equation}
\sigma_w \simeq 16m_a f_a^2\ .
\end{equation}
Using the energy density of long strings (\ref{eq:rhols}),
the density parameter due to the contribution of the axions
produced by infinitely long axionic string decay is rewritten as
\begin{equation}
\Omega_a({\rm string}, \infty) h^2 = 1.2\times 10^{-4}\Delta
\left(\frac{\zeta}{13}\right)\ ,
\end{equation}
where the common normalization factor to all $\Omega_a$ is defined by
\begin{equation}
\Delta\equiv \left(\frac{f_a}{10^{9}{\rm GeV}}\right)^{1.175}
\left(\frac{\cal N}{72.25}\right)^{-1.16}
\left(\frac{\Lambda_{\rm QCD}}{200{\rm MeV}}\right)^{-0.65}
\left(\frac{m_a^0}{6.2\times10^{-3}{\rm eV}}\right)^{0.825}\ .
\end{equation}
Note that $\Omega_a({\rm string}, \infty)$ is independent of $\kappa$
since the long string distribution is determined only by $\zeta$.
On the other hand, using the energy density of loops (\ref{eq:rholp}),
the density parameter of the axions by the loop decay source is also
rewritten as
\begin{equation}
\Omega_a({\rm string}, {\rm loop}) h^2 = 1.1\times 10^{-3}
\Delta \left(\frac{\alpha}{\kappa}\right)^{3/2}\left[1-\left(1+
\frac{\alpha}{\kappa}\right)^{-3/2}\right]\left(\frac{C_a}{0.4}\right)
\left(\frac{\Gamma_a}{65}\right)\left(\frac{\zeta}{13}\right)\ .
\label{eq:ostloop}
\end{equation}
The ratio of $\Omega_a({\rm string}, {\rm loop})$ to
$\Omega_a({\rm string}, \infty)$ is plotted in Fig. \ref{fig:ostring}
with the conventional values of the parameter
such as $\alpha =(0.1\sim 1)\kappa$.\cite{VSa}
Evidently the contribution from the loops is dominant
in the range of reasonable parameter values.

The energy density of the axions from the axionic domain wall
when they become non-relativistic can be calculated using that
at $t_1$, $\rho_w(t_1)$, as
\begin{equation}
\rho_a\left(t\right)=\rho_w\left(t_1\right)\left(\frac{a\left(t_1\right)}
{a\left(t\right)}\right)^3\left(\frac{\langle E_a \rangle}
{m_a}\right)^{-1}\ ,
\end{equation}
where $a$ is the scale factor and $\langle E_a \rangle$ is
the averaged energy of the emitted axions from the walls at $t_1$.
We have obtained the relation
\begin{equation}
\frac{\langle E_a \rangle}{m_a}\simeq 3
\end{equation}
in simulations of the axionic domain walls.\cite{NK} Thus
$\Omega_a({\rm wall})$ will be estimated using this numerical value.

\begin{figure}
\epsfile{file=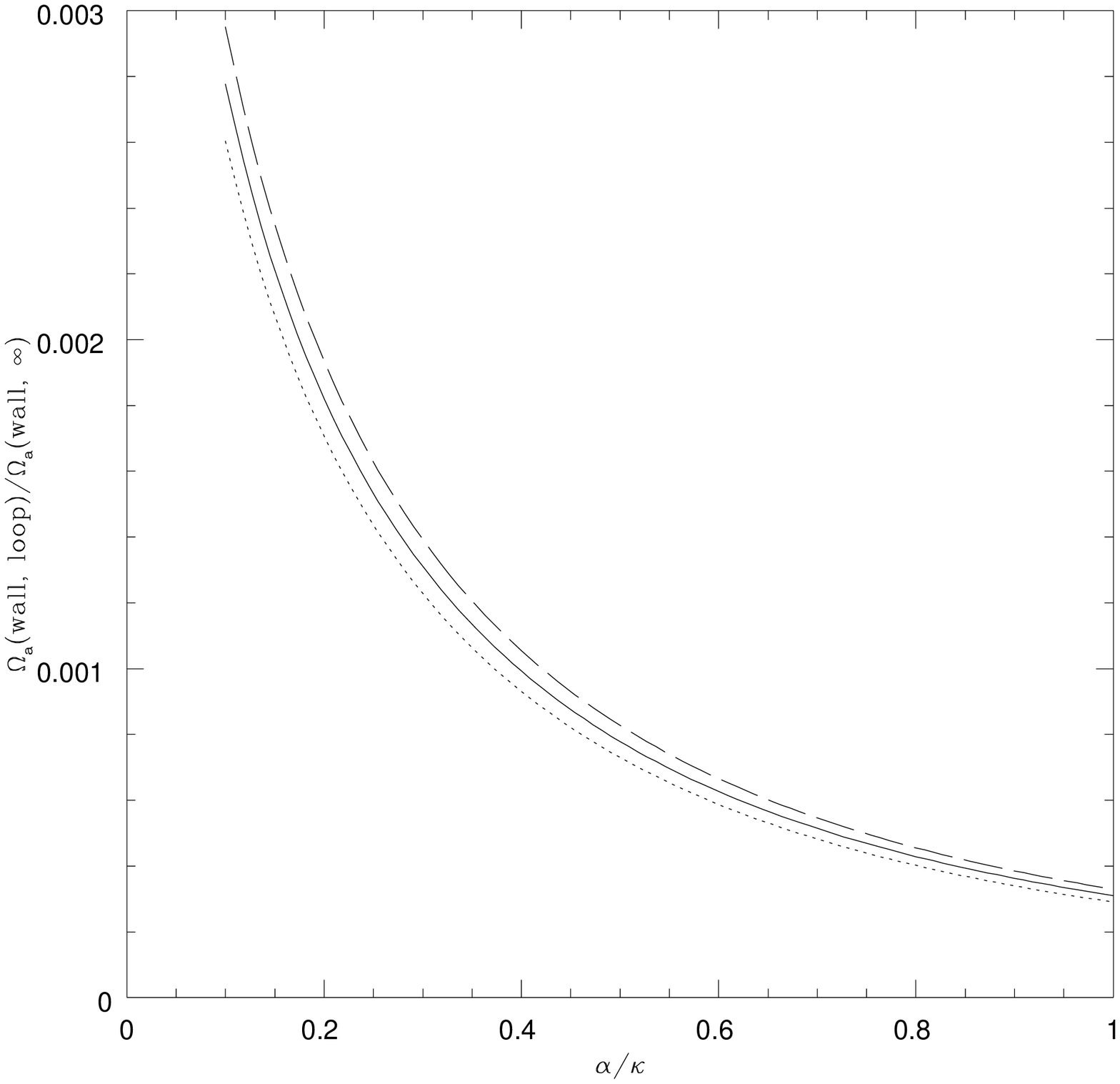,width=15.0cm}
\caption{The ratio of the energy density of the axions emitted from
the axionic domain walls surrounded by the string loops to that
attached to the long strings is plotted against $\alpha/\kappa$.
The dotted line corresponds to the case that $\kappa =0.15$, the solid
line to $\kappa=0.16$, and the dashed line to $\kappa=0.17$.}
\label{fig:owall}
\end{figure}

We can rewrite the formula (\ref{eq:rhols}) as
\begin{equation}
\rho_s^\infty\left(t_1\right)=\mu_s \frac{1}{L_1^2} \ ,
\end{equation}
where $L_1\equiv t_1/\sqrt{\zeta}$ is the characteristic length scale
of the string network, so that we can easily regard this as being
equal to the averaged distance between the axionic domain walls and
at the same time, their mean radius. Thus the energy density of the
axionic domain walls attached to the long axionic strings is derived as
\begin{equation}
\rho_w^\infty\left(t_1\right)=\frac{\sigma_w L_1^2}{L_1^3}
=\sigma_w \sqrt{\zeta}\frac{1}{t_1}\ ,
\end{equation}
and the axion density due to these walls is estimated to be
\begin{equation}
\Omega_a({\rm wall}, \infty) h^2 = 4.6\times 10^{-3}
\Delta\left(\frac{\zeta}{13}\right)^{1/2}\ .
\label{eq:owlong}
\end{equation}
We can expect that by a string loop of length $l$, a piece of wall
of area $l^2/4\pi$ surrounded by the incident loop, will be created.
Then the energy density of the axions due to the axionic domain walls
which originate from the loops is written as
\begin{equation}
\rho_w^{\rm loop}\left(t_1\right)=\int_0^{\alpha t_1} \sigma_w
\frac{l^2}{4\pi}\frac{\rho_s^{\rm loop}\left(t_1\right)}{\mu_s l}dl\ ,
\end{equation}
and the contribution to the present density parameter from these
axions is written as
\begin{eqnarray}
& & \Omega_a({\rm wall}, {\rm loop}) h^2=1.0\times 10^{-4}\Delta
\left(\frac{C_a}{0.4}\right)\left(\frac{\zeta}{13}\right)\nonumber \\
& & \times \sqrt{\alpha \kappa}\left[2\left\{\left(1+\frac{\kappa}{\alpha}
\right)^2 +2\left(1+\frac{\kappa}{\alpha}\right)-\frac{1}{3}\right\}
\left(1+\frac{\kappa}{\alpha}\right)^{-3/2}-\frac{16}{3}\right]\ .
\end{eqnarray}
The results indicate that $\Omega_a({\rm wall}, \infty)$ is greater
than $\Omega_a({\rm wall}, {\rm loop})$ for the region of reasonable
parameter values in contrast to the axionic string case in which
the loop contribution is more significant. We can see this property
in Fig. \ref{fig:owall} more clearly. This is because that the larger
a defect grows, the more dominant the wall energy becomes in relation
to the string energy. The dependence on $\kappa$ implies that
when $\alpha/\kappa$ is constant, a larger value of $\kappa$ and also
of $\alpha$ is favorable for the greater share of the walls
attached to loops. This can be interpreted as the effect that
the larger loop creation size must be important
since the wall piece distribution at a certain time $t_1$
is in question in the wall case. The distinct feature compared with
the string contribution in Fig. \ref{fig:ostring} is that
$\Omega_a({\rm wall}, {\rm loop})/\Omega_a({\rm wall}, \infty)$
becomes smaller as $\alpha/\kappa$ increases. This can be explained,
roughly speaking, by noting that the energy density of walls is
a remnant of the axionic string energy, so a decrease in the
efficiency of the axion emission from the strings implies an increase
in the amount of energy possessed by the axionic domain walls.

\begin{figure}
\epsfile{file=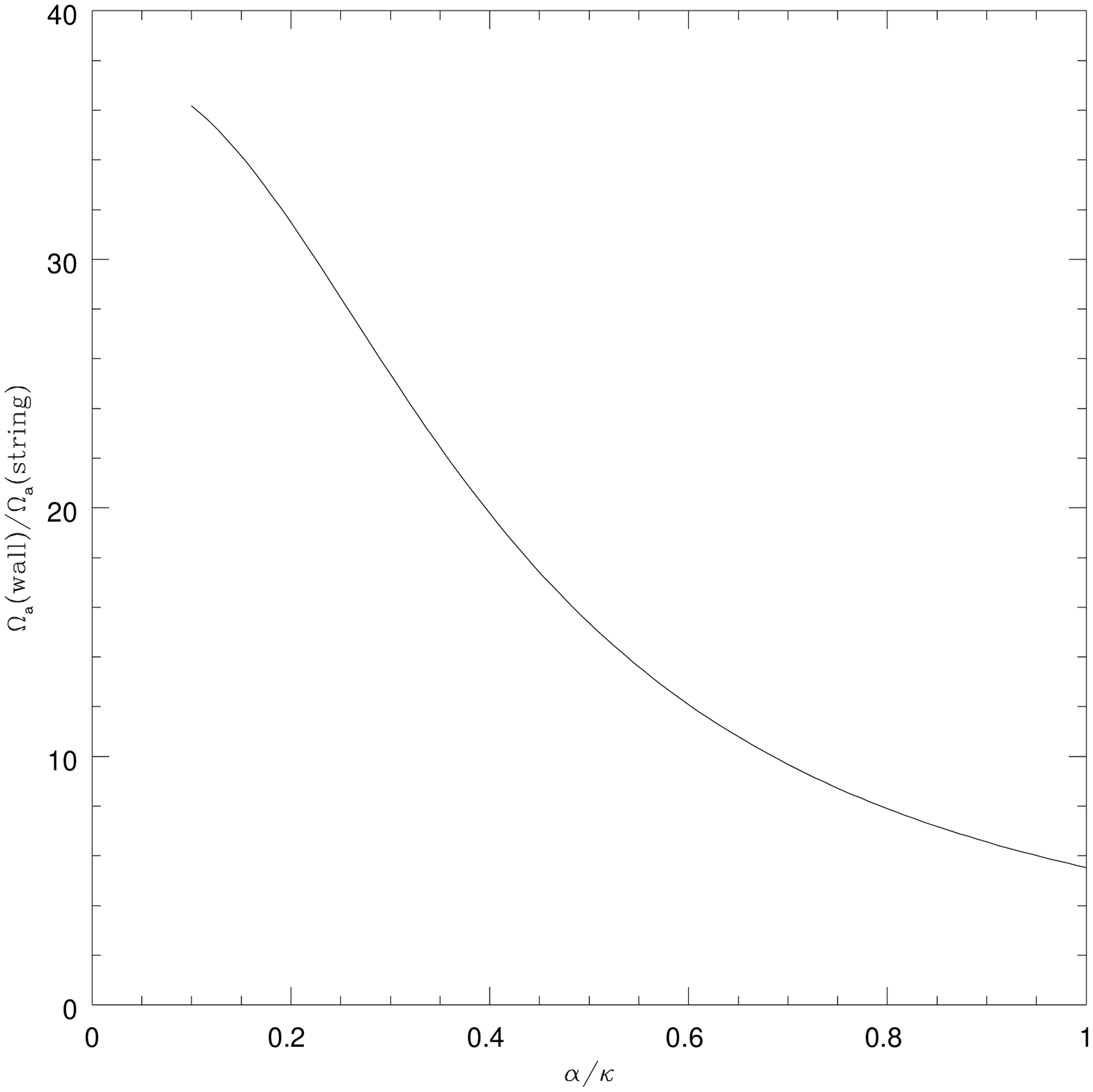,width=15.0cm}
\caption{The ratio of the energy density of the axions produced by
the axionic domain walls to that by the axionic strings versus
$\alpha/\kappa$ is shown in the case when $\kappa=0.16$.}
\label{fig:oratio}
\end{figure}

Finally we compare $\Omega_a({\rm wall})=\Omega_a({\rm wall}, \infty)
+\Omega_a({\rm wall}, {\rm loop})$ with $\Omega_a({\rm string})$ \\
$=\Omega_a({\rm string}, \infty)+\Omega_a({\rm string}, {\rm loop})$,
which is demonstrated in Fig. \ref{fig:oratio}.
The ratio takes the form
\begin{equation}
\frac{\Omega_a({\rm wall})}{\Omega_a({\rm string})}= \frac{4.6}
{0.12+1.1\left(\frac{\alpha}{\kappa}\right)^{3/2}\left[1-\left(1+
\frac{\alpha}{\kappa}\right)^{-3/2}\right]\left(\frac{C}{0.4}\right)
\left(\frac{\Gamma_a}{65}\right)}\left(\frac{\zeta}{13}\right)^{-1/2}\ ,
\label{eq:oratio}
\end{equation}
and it satisfies
\begin{equation}
\frac{\Omega_a({\rm wall})}{\Omega_a({\rm string})}\approx 5\sim 36\ ,
\end{equation}
which means the wall contribution is larger than or comparable to
that of the strings. Although Fig. \ref{fig:oratio} shows
the case when $\kappa=0.16$, we cannot distinguish its appearance
from the other parameter value cases, such as $\kappa=0.15$ or
$\kappa=0.17$. This is because the dominant contribution from
axionic domain walls is that from the walls of long strings which
is independent of $\kappa$, as can be seen from Eq. (\ref{eq:owlong}),
and the string contribution also depends only on $\alpha/\kappa$,
which can be seen from Eq. (\ref{eq:ostloop}).

It should be noted that all the energy density contribution is
in proportion to the square root of the estimation time :
\begin{equation}
\Omega_a \propto t_1^{1/2}\ ,
\end{equation}
so that the total amount of the axion would be twice as large as
the value obtained in this paper if we were to calculate it at $t_w$,
which is employed in Ref. \cite{BS}, since $t_w \sim 4 t_1$.
The relative fraction of the axionic domain wall contribution,
however, is invariant even when the time of the final string
expiration and the wall collapse would be shifted. We may say that
the energy density of the axions produced by the axionic domain walls
which stem from the long axionic strings yields the most dominant
contribution to the present axion density among the axion production
mechanisms by the topological defects.

\section{Conclusions}

We have investigated whether the scale-invariant distribution
of the axionic strings is realized. This is very important in
the estimation of the axion production by axionic string
radiation. As the key element in the confirmation or refutation of
this assumption, the scattering cross section of the relativistic
fermions from the axionic string has been calculated using
a simple model with a generic form of interaction.

The one-scale model conjecture can be justified when the Peccei-Quinn
charges of the particles are properly chosen so that the friction
regime of the axionic strings ends up at an energy scale which is
much higher than that when the axionic domain wall rules the evolution
of the wall-string system. For instance, the hadronic axion has
only slight interactions with the leptons. Then the scatter of
these particles can be neglected, which means the correction
factor becomes effectively small, and $T_*$ increases. Even if
the parameter $g_f$ has an amplitude of order 1, which is
the most natural choice in various models, the era near $T_1$
when the main contribution from the axions of the axionic
string origin is radiated does not seem to be in the friction regime.

Since the scaling distribution of the axionic string is shown to
be more plausible, the axion energy density due to the axionic domain
walls can be calculated using the wall distribution inferred from it.
Consequently, we have found that the wall contribution,
$\Omega_a({\rm wall})$, is not less important than
$\Omega_a({\rm string})$ within the region of reasonable values of
the model parameter and the estimation time. The relative importance
of the walls stuck to the infinitely long strings reflects the property
of a wall that it is a two-dimensional defect and its energy is
proportional to the area surrounded by the string.

Note that the ratio (\ref{eq:oratio}) is proportional to
$\zeta^{-1/2}$ and the numerical value, $\zeta =13$, is evaluated
by simulations of gauged strings. In the case of global axionic
strings, a long-range interaction would make the smoothing
scale larger, i.e., $\zeta$ should be much smaller.
Then the ratio of $\Omega_a({\rm wall})$ to $\Omega_a({\rm string})$
grows greater, although their absolute amplitudes decrease.

If the line energy density of the axionic string, $\mu_s$, could be
reduced, the friction dominant era of the axionic strings would
continue so long that the string distribution around $T_1$ might
be modified as the relation (\ref{eq:tpropo}) indicates. In such
a situation, the energy density of the loop tends to be small.
The dominance of $\Omega_a({\rm wall})$ over the string contribution
is, however, preserved as long as a certain quantity of the horizon
scale string remains.

In any case, the evolution of the axionic strings and their manner of
distribution should be analyzed in detail. Since the axionic strings
are global strings, they have wide extended line energy density,
and this may affect the estimation of $\Omega_a({\rm wall})$ and
$\Omega_a({\rm string})$ in addition to the enhanced smoothing.
Even if the scaling distribution is correct, the actual calculation
of $\Omega_a$ is indispensable in the evaluation of the parameters
such as the minimal loop-size and the energy loss rate by
the gravitational waves, which are concerned with the ambiguity
of the typical size and the number density of the strings.
Moreover, the true energy spectrum of radiated axions is also closely
connected with the string dynamics. Further investigation of
axionic strings must be made carefully in order to provide
precise constraints on the axion model.

\section*{Acknowledgments}

The author acknowledges support from a JSPS fellowship.
This work was partially supported by the Japanese Grant
in Aid for Scientific Research Fund of the Ministry of Education,
Science, Sports and Culture No. 5110.

\end{document}